\documentclass[a4paper]{article}

\usepackage{INTERSPEECH2018}
\usepackage{graphicx}
\usepackage[parfill]{parskip}
\usepackage{tipa}
\usepackage{phonrule}
\usepackage{amsmath, amscd, amsthm, amssymb, mathrsfs,amsfonts}

\title{Towards Automatic Speech Identification from Vocal Tract Shape Dynamics in Real-time MRI}
\name{Pramit Saha$^1{^*}$\thanks{* indicates equal contribution}, Praneeth Srungarapu$^1{^*}$, Sidney Fels$^1$}
\address{
  $^1$ Department of Electrical and Computer Engineering, University of British Columbia}
\email{pramit@ece.ubc.ca, praneethsv@ece.ubc.ca, ssfels@ece.ubc.ca }

\begin{document}

\maketitle
\begin{abstract}
  Vocal tract configurations play a vital role in generating distinguishable speech sounds, by modulating the airflow and creating different resonant cavities in speech production. They contain abundant information that can be utilized to better understand the underlying speech production mechanism. As a step towards automatic mapping of vocal tract shape geometry to acoustics, this paper employs effective video action recognition techniques, like Long-term Recurrent Convolutional Networks (LRCN) models, to identify different vowel-consonant-vowel (VCV) sequences from dynamic shaping of the vocal tract. Such a model typically combines a CNN based deep hierarchical visual feature extractor with Recurrent Networks, that ideally makes the network spatio-temporally deep enough to learn the sequential dynamics of a short video clip for video classification tasks. We use a database consisting of 2D real-time MRI of vocal tract shaping during VCV utterances by 17 speakers. The comparative performances of this class of algorithms under various parameter settings and for various classification tasks are discussed. Interestingly, the results show a marked difference in the model performance in the context of speech classification with respect to generic sequence or video classification tasks. 
\end{abstract}
\section{Introduction}
The vocal tract~\cite{veena2016study} is the most important component of speech production that can be divided into two distinctive parts - the \textit{oral tract} (ranging from larynx to lips) and the \textit{nasal passage}, coupled to the oral cavity through velum. During air expulsion, this tubular passageway modifies its position and shape (length and cross sections) in such a way that results in production of various sounds and their acoustic representations.

A long-standing issue in articulatory speech research concerns the estimation of vocal tract configuration and its mapping with the acoustic parameters~\cite{mitra2017joint}. This is extremely challenging, not only because the vocal tract is difficult to access, but also due to its complex biological structure and the rapid movement of its articulators. The speech production process is essentially “non-stationary” and generally, it is the rapid transition between different articulatory states that generates the speech sounds. Hence, extraction of the dynamic information of vocal tract geometry is crucial for identifying and synthesizing speech.

The relationship between the articulatory gestures and acoustic speech characteristic has been well explored in terms of either \textit{articulatory-to-acoustic mapping} which deals with producing audio speech signal from vocal tract configurations or the more common \textit{acoustic-to-articulatory inversion}, which aims at recovering the vocal tract shapes from the speech~\cite{hueber2011statistical}. 

In this paper, our primary goal is to investigate the end-to-end mapping of the forward pathway, \textit{i.e} learning to identify speech sounds corresponding to subject-invariant vocal tract geometry configurations. There are two main ways of achieving this: the \textit{modeling} approach or the \textit{corpus-based} approach~\cite{hueber2011statistical}. Research in the former approach mostly focus on three basic steps - \textit{articulatory model representation}, generally achieved by speaker-specific \textit{geometric modeling} of articulator shapes and positions; speaker-invariant \textit{statistical modeling} of standard articulator representations; or neuro-muscular controlled \textit{biomechanical modeling} of articulator motions, followed by 
\textit{acoustic transfer function computation} step, and time or frequency based
\textit{acoustic characterization}.

Conversely, the \textit{corpus-based} methodology utilizes parallel articulatory-acoustic database consisting of speech sounds recorded with simultaneous articulator movements, through \textit{code-book} or  \textit{statistical} (discriminative or generative) techniques. While the \textit{code-book} approach is a direct way to retrieve the nearest articulatory configurations, corresponding to given \textit{`query'} configuration from the database and produce an audio spectrum interpolated from available acoustic representations, the \textit{statistical} model utilizes supervised machine learning techniques to learn the forward connection.

But, all the available methods rely solely on the articulatory time-series positional data from a few specified points on tongue, lip and jaw acquired using electromagnetic articulography, or sometimes combined with electropalatography and laryngograph. While it is considerably easier and computationally less expensive to work with a set of tracked input points at a handful of sensor locations within the anterior part of the tract, it misses significant articulatory information like velar and pharyngeal constrictions, and others, thereby neglecting the spatial information of the complex vocal tract geometry. Dynamic vocal tract-imaging data, on the other hand, effectively records the labial, lingual and jaw motion along with the articulation of the velum, pharynx and larynx, and structures like the upper palate, pharyngeal wall \textit{etc} which cannot be captured by the above-mentioned techniques. Besides, such imaging data helps to comprehend the generation of coronal, pharyngeal and nasal segments and to visualize the inter-articulator coordination for generation of multi-gestural segments in speech. Hence, there is an inherent urge to explore the informations available from medical imaging modalities on the articulatory-acoustic forward pathway utilizing dynamic information of the entire upper airway. 

However, it is extremely challenging and computationally expensive to achieve a direct end-to-end mapping of the entire vocal tract geometry acquired by real time imaging modalities to the corresponding speech sound output. Recently, real-time magnetic resonance imaging (rtMRI) has emerged as a powerful tool to acquire the complex spatio-temporal visual information about the dynamic shaping of the vocal tract during speech production~\cite{toutios2016advances}. This has opened a new pathway to investigate the underlying articulatory geometric representation of the speech signals in real-time, thereby developing a better understanding of the geometry-to-acoustic mapping. However, the poor spatio-temporal resolution as well as the presence of a substantial number of artifacts and noise and limited data make automatic extraction and interpretation of features an extremely difficult problem~\cite{lingala2016recommendations}. 

To the best of our knowledge, this work is the first discriminative approach that attempts to achieve a forward mapping of vocal tract shape dynamics from rtMRI to the speech output. In order to exploit the dynamics of open and closed vocal tract configurations along with the rapid articulator movements, we have selected 51 VCV transition tokens as the target classes. We employ a deep learning strategy to extract the per-frame spatial features and simultaneous inter-frame temporal features. The primary goal of this paper is to investigate the performance of a promising visual action detection and classification algorithm named Long term Recurrent Convolutional Neural Networks (LRCN)~\cite{donahue2015long}, when applied for speech identification tasks from rtMRI videos. Though the technique fails to provide satisfactory accuracy for the VCV sequence identification task, it shows promising results for classifying the individual vowels and acceptable performance for consonants. We discuss the issues that make the targeted speech identification task different from conventional video-classification~\cite{karpathy2014large} or action recognition tasks~\cite{simonyan2014two} and provide insights regarding possible ways of improvement.

\section{Illustration of the Proposed Methodology}
\subsection{Long-term Recurrent Convolutional Networks model}
\begin{figure}
	\centering
	\includegraphics[width=8cm, height=6cm, keepaspectratio,]					{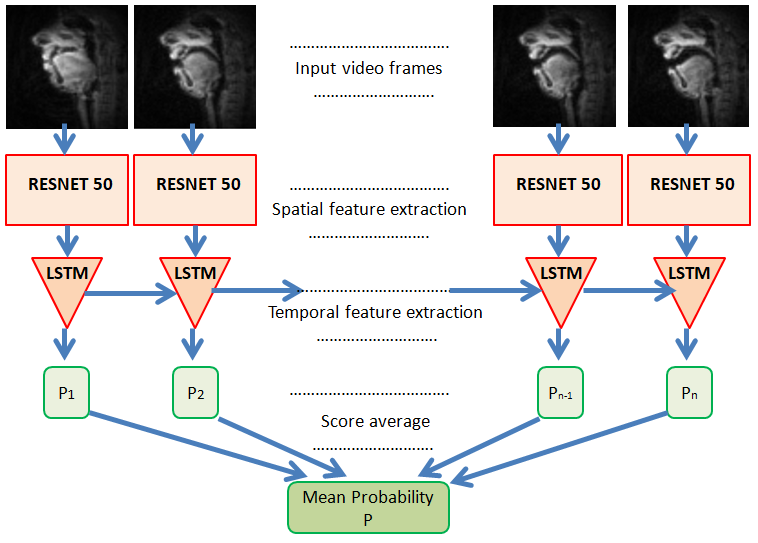}
	\caption{Overview of the LRCN model}
	
\end{figure}
\subsubsection{Background}
Among the most widely used algorithms for video based action recognition, 3D Convolutional Networks~\cite{tran2015learning}, Two stream Convolutional Networks~\cite{simonyan2014two} and Long-term Recurrent Convolutional Networks (LRCN)~\cite{donahue2015long} deserve special mention in our context. This class of algorithms incorporate spatial and temporal feature extraction steps to capture complementary information from the individual still frames as well as between the frames, which is associated to our primary goal in this work. However, the two stream architecture involves CNN trained on multi-frame dense optical flow images for the temporal recognition stream. Since the dataset we are using~\cite{sorensen2017database}, has a significant amount of non-stationary noise and artifacts, such algorithm gives more emphasis on the noise instead of the vocal tract shape change, which in turn leads to erroneous sound classification. On the other hand, though 3D ConvNets show faster performance on our dataset, the temporal pooling layers are too weak to extract long term temporal features spanning over a duration of 3 seconds. The end-to-end trainable LRCN networks combine CNN based deep visual feature extractor with LSTM based long-term temporal dynamics extractor and are shown to have sufficient recurrence of the latent variables over temporal domain.  Thus, we utilize the LRCN model to investigate whether the speech recognition problem can be viewed similar to a video action classification/recognition task, with the vocal tract movement resembling the action and the VCV sequences as the final mapped output interpreted from the action. However, the performance of the other two is not significant enough in the current context and hence we decide to confine our discussion with LRCN.
\subsubsection{Convolutional Neural Network (CNN)}
CNN architectures~\cite{krizhevsky2012imagenet} are composed of sequence of layers - Convolutional Layers (Learnable spatial filters), Pooling Layers (Downsamplers) and Fully Connected Layers (Dense layers with full connections), along with linear/non-linear activations. We use substantially deep residual learning framework, also known as ResNet~\cite{he2016deep}, composed of series of Identity and Convolutional blocks in this work. 
\begin{figure*}[tb]
\centering
\includegraphics[width=\textwidth, height=1.8cm]{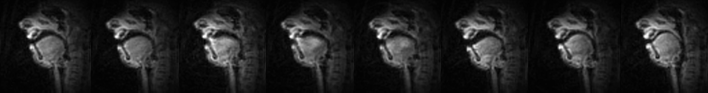}
\caption{Frames of rtMRI videos for speaker F1 producing [asa]. Time progresses from left to right.}
\end{figure*}
\subsubsection{Recurrent Neural Network (RNN)}
Recurrent neural Networks are capable of exhibiting dynamic temporal behavior~\cite{graves2013speech}. LRCN network~\cite{donahue2015long} uses RNN composed of LSTM units~\cite{sak2014long} which are capable of learning complex temporal dynamics by mapping the input vectors to the output vectors through a sequence of hidden steps defined by recurrence relations. They are effective in solving the exploding and vanishing gradient problems associated with conventional RNN networks~\cite{hochreiter1997long}. The LSTM module utilized here~\cite{zaremba2014learning}, consists of a memory cell, an input gate, an output gate, a forget gate and an input modulation gate. Such networks are able to explore temporal behavior due to inter-layer directed units, as shown in Fig 1.
\subsection{Vowel-Consonant-Vowel Identification through LRCN}
We consider the VCV identification from the rtMRI as the `\textit{sequential input to fixed output}' problem with videos of arbitrary length $T$ as input and prediction of single labels corresponding to each video, as the output. The video is split into $T$ individual frames or image sequences and fed to $T$ convolutional network layer, each network layer implying a complete ResNet architecture and then connected through two fully connected layers with $2048$ and $1024$ neurons with $0.5$ drop-outs respectively to an $N$ layered LSTM with $M$ hidden nodes. Next, the LSTM predicts the probabilities of speech output class for each of the time frames and are averaged to yield the final class probability score across the entire sequence demonstrated in Fig 1.

\section{Experiments and Results}
\subsection{Dataset preparation}
We evaluate our architecture on the USC Speech and Vocal Tract Morphology MRI Database~\cite{sorensen2017database} which includes $2D$ real-time MRI of vocal tract shaping of $17$ speakers ($9$ female and $8$ male) along with simultaneous denoised audio recording. This database provides the resource to relate rtMRI data capturing dynamic vocal tract shapes and speech variability, thereby helping our attempt to explore the forward mapping pathway from imaging data. The imaging sequence has a \textit{frame rate} of 23.18 frames/second, \textit{slice thickness} of $5 mm$, \textit{spatial resolution} of $2.9 mm^2$/pixel and \textit{field of view} of $200mm \times 200mm$. Further details regarding the database are available in~\cite{narayanan2014real}. 

The database contains 3 repetitions corresponding to each speaker for each of 51 VCV utterances (\textit{\textipa{apa}, \textipa{upu}, \textipa{ipi}, \textipa{ata}, \textipa{utu}, \textipa{iti}, \textipa{aka}, \textipa{uku}, \textipa{iki}, \textipa{aba}, \textipa{ubu}, \textipa{ibi}, \textipa{ada}, \textipa{udu}, \textipa{idi}, \textipa{aga}, \textipa{ugu}, \textipa{igi}, \textipa{a\textsc{T}a}, \textipa{ithi}, \textipa{uthu}, \textipa{asa},
\textipa{usu}, \textipa{isi}, \textipa{aSa}, \textipa{uSu}, \textipa{iSi}, \textipa{ama},
\textipa{umu}, \textipa{imi}, \textipa{ana}, \textipa{unu},
\textipa{ini}, \textipa{ala}, \textipa{ulu}, \textipa{ili}, \textipa{afa}, \textipa{ufu}, \textipa{ifi}, \textipa{a\reflectbox{\textscl}a}, \textipa{u\reflectbox{\textscl}u}, \textipa{i\reflectbox{\textscl}i},
 \textipa{aha}, \textipa{uhu}, \textipa{ihi}, \textipa{awa}, \textipa{uwu}, \textipa{iwi}, \textipa{aja}, \textipa{uju}, \textipa{iji}}). We pre-segment these into a total of 2754 videos, each containing the entire length of single VCV utterance. We further divide the total available data into training dataset having 2268 videos  for 14 speakers and testing dataset with the remaining 486 videos. The testing dataset containing videos corresponding to 3 speakers with 3 repetitions were absolutely unseen during the training phase. Eight frames of a sample test data, where a female speaker F1 produces the VCV token \textit{[asa]} has been shown in Fig 2.

\subsection{Training}
Sixteen image frames are extracted from each of the videos with a stride of 3, excluding the silent frames. The target is to classify the videos into any of the 51 VCV labels. For the CNN part, we utilize the 50 layered Residual Network (ResNet50) that is pre-trained on the ILSVRC-2012 dataset~\cite{russakovsky2015imagenet}. This speeds up the training process with the optimized weights and further restricts overfitting in the current small dataset. We vary the parameters one by one keeping all others fixed and simultaneously conduct evaluations for various combinations to select the set of optimum values. The batch size is varied from 32 to 128 and set to the optimum value of 64. Similarly, the number of steps/epoch for the training is changed from 100 to 1000 and fixed at 500, which gives better accuracy. The range of values tried for layers and nodes of LSTM network are 1 to 5 and 32 to 512 respectively. We conclude that the respective optimal values are 1 and 256. The total number of epochs is kept at 140.  The drop out value for the LSTM network is varied from .5 to .97 and finally fixed at $0.9$. The performance considerably decreases when any activation function except softmax is used.

With the optimal value of the set of parameters remaining fixed, the number of target classes were changed from 51 to 3 and 17 for individual vowels and consonants, respectively. Subsequently, the target class labels were also modified.

\subsection{Performance Evaluation}

We conducted evaluation for various combinations of parameters and classes (V, C and VCV).
In the objective evaluation, the loss measure was given by categorical cross-entropy and various metrics such as Top-1, Top-5 and Top-10 categorical accuracy were employed to analyze its performance. 
 
From Table 1, it can be observed that the mapping accuracy markedly improves from VCV to vowels. The algorithm achieves highest performance of $0.96$ when we classify the data into 3 vowel classes, as shown in Fig 3. The performance unexpectedly drops to $0.68$ in terms of Top-1 accuracy for 17 consonant identifications. A careful analysis of the predicted classes shows that, the videos are wrongly classified because of their spatial characteristics rather than the temporal features. In other words, it tends to shift towards speaker based classification, rather than the targeted consonant based identification. This is because the number of speakers is also 17 and hence, in this case, the ResNet architecture dominates the trained LSTM layer which makes the identification task much more difficult than vowel identification. 

\begin{figure}
	\centering
	\includegraphics[width=8cm, height=8cm, keepaspectratio,]					{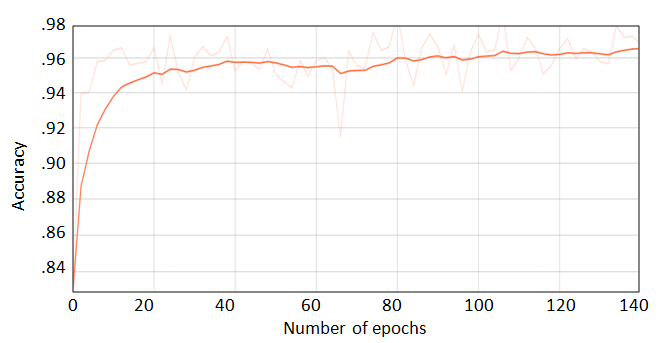}
	\caption{The Top-1 accuracy for vowel identification}
\end{figure}

Top-1 classification accuracy for test VCV sequences is $0.42$ as shown in Fig 4. Top-5 and Top-10 accuracies reach $0.76$ and $0.93$ respectively. Though it has demonstrated Top-1 accuracy of .8 for video action recognitions, it fails to maintain the accuracy for the current sound identification task. The degradation of accuracy may be due to excessive increase in number of controllable parameters with increase in number of classes, which would require more training data for estimation.

To investigate this issue further, we divided the dataset into 3 parts with 17 target classes, to avoid ambiguity among tokens involving similar articulator movements. As a consequence, the results increased by a considerable margin of $0.12$ and the maximum accuracy reached around $0.55$ which proves that the similar articulatory movements getting mapped to different sounds is a vital issue that the LRCN algorithm is unable to detect.


\section{Discussion and Limitation}
\begin{table}[t]
  \caption{Speech identification performance}
  \label{tab:word_styles}
  \centering
  \begin{tabular}{ll}
    \toprule
    \textbf{Identification task}      & \textbf{Top-1 accuracy}            \\
    \midrule
    Vowel                    & $0.96$                                       \\
    Consonant                   & $0.68$                               \\
    VCV              & $0.42$                          \\ 
    
    \bottomrule
  \end{tabular}
\end{table}
Automatic speech recognition from the movements and deformations of vocal tract articulators as visualized from an imaging modality is an incredibly challenging task. 
The first issue associated with this, is the inter-speaker anatomical differences in vocal tract. This is not a considerable problem in EMA, as the EMA recordings are more concerned with tracking the time varying trajectories of selected points on the tongue and palate. However, while extracting features from imaging modalities, these anatomical differences restrict a particular model from generalizing the structural features like the shapes and lengths of the articulators, which play secondary role in speech identification. Hence, it is difficult to identify the articulator positions and contours due to this inter-subject variability. This makes the conventional methods like Hidden Markov Models~\cite{hiroya2004estimation}, Maximum Likelihood Mapping Models~\cite{hodgen2001stochastic}, Gaussian Mixture Models~\cite{toda2008statistical} quite ineffective in extracting appropriate sequential, structural features from the imaging dataset. 

Secondly, the subtle and rapid changes of the tract physiology makes the speech recognition task more difficult. This is conceptually different from the conventional action recognition tasks, where the target is to classify the video action space based on widely varying user-movements. In our case, a speaker-specific frame-by-frame analysis of the image sequences corresponding to various VCV tokens demonstrate nearly similar tongue motion and vocal tract contour variation. This makes it difficult to detect significant changes in articulator movements for the consonant part of the VCV utterances. Also, the action recognition tasks involve an underlying spatial object detection task to interpret the actions from the video-clips, through spatial networks. However, such networks~\cite{donahue2015long,simonyan2014two,tran2015learning} fail to extract meaningful features from rtMRI frames to map them to distinct articulatory movements towards speech identification. 
\begin{figure}
	\includegraphics[width=8cm, height=6cm, keepaspectratio,]					{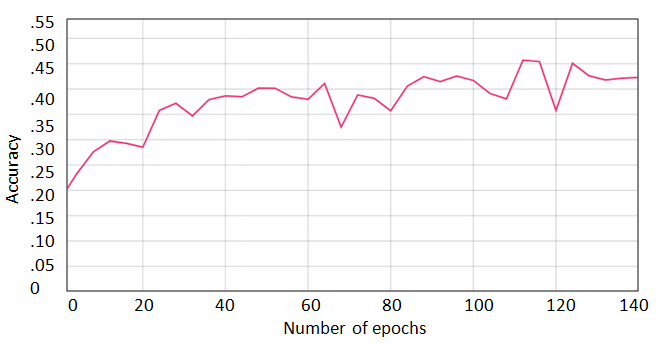}
	\caption{The Top-1 accuracy for VCV identification}
	
\end{figure}

Thirdly, more than two-third of the total time duration for each video clip is occupied for uttering the two `V's and the remaining time is utilized for the intermediate `C'. So there are more frames corresponding to the vowels than the consonants, which means more feature extraction from those frames and heavier emphasis on the vowels in the classification task. Besides, the training data corresponding to each vowel class is more than that corresponding to each consonant or each transition. The availability of training data is seen to have an enormous influence on the accuracy of the methods. Thus, while the algorithm shows great potential for vowel identification from the tokens, the accuracy reduces for consonants and gets further decreased for the entire VCV identifications. 

Fourthly, rtMRI images have low spatial and temporal resolution and are infested with various noises and reconstruction artifacts~\cite{lingala2016recommendations}. This work mostly focuses on addressing the speech identification issue from the raw images and hence, we did not incorporate any preprocessing module. However, it might be interesting to investigate how noise and artifact reduction or resolution enhancement effects the performance accuracy.

Lastly, we observed that the acoustic pitch and energy level of the audio signals exhibit significant variations for several VCV tokens, which could be utilized to achieve better classification. One way to incorporate this is to associate the target class labels with such acoustic variables and take advantage of multi-variate loss function towards accurate prediction scores.

\section{Conclusion and Future Directions}
A promising deep learning based video action classification technique named Long Term Recurrent Convolutional Network (LRCN) has been trained and tested on 2754 videos with 51 VCV tokens. 
Results showed that the targeted articulatory-to-acoustic mapping was satisfactorily approximated for vowels and acceptably for consonants. However, the classification performance demonstrated a considerable decrease in accuracy, while identifying the entire VCV transitions, which implies that the model alone is not sufficient to distinguish these transitional tokens. 

Along with the structural vocal tract model, it might be interesting to explore whether augmentation from the vocal tract flow model~\cite{zappi2016towards} can assist the algorithm to differentiate between consonants that involve similar articulator movements. In future, we will investigate this issue as well as address whether its performance can be increased by coupling it with an underlying biomechanical model for speech production~\cite{vogt2006efficient,gick2014speech,stavness2012automatic} or using Deep belief Networks (DBN)~\cite{wei2016mapping}.

\section{Acknowledgements}


This work was funded by the Natural Sciences and Engineering Research Council (NSERC) of Canada and Canadian Institutes for Health Research (CIHR).
The authors would like to thank Amir Hossein Abdi and C. Antonio Sanchez of HCT Lab for providing their valuable insights. 
\bibliographystyle{IEEEtran}

\bibliography{mybib}


\end{document}